\documentclass[reqno,10pt]{amsart}
\usepackage[english]{babel}
\usepackage{amsthm}
\usepackage{amsmath}
\usepackage{amssymb}
\usepackage[latin1]{inputenc}

\makeatletter

\@addtoreset{equation}{section} \makeatother
\newtheorem{theorem}{Theorem}[section]
\newtheorem{remark}[theorem]{Remark}
\newtheorem{lemma}[theorem]{Lemma}
\newtheorem{proposition}[theorem]{Proposition}
\newtheorem{corollary}[theorem]{Corollary}
\newtheorem{definition}[theorem]{Definition}

\newcommand{\be}{\begin{equation}}
\newcommand{\ee}{\end{equation}}

\newcommand{\N}{{\mathbb N}}

\newcommand{\R}{{\mathbb R}}
\newcommand{\LL}{{\mathbb L}}

\newcommand{\noi}{\noindent}

\newcommand{\ben}{\begin{enumerate}}
\newcommand{\een}{\end{enumerate}}
\newcommand{\bit}{\begin{itemize}}
\newcommand{\eit}{\end{itemize}}
\newcommand{\edoc}{\end{document}}

\newcommand{\bdefi}{\begin{definition}}
\newcommand{\btheo}{\begin{theorem}}
\newcommand{\bprop}{\begin{proposition}}
\newcommand{\brema}{\begin{remark}}
\newcommand{\bcoro}{\begin{corollary}}
\newcommand{\blemm}{\begin{lemma}}
\newcommand{\bexam}{\begin{example}}

\newcommand{\edefi}{\end{definition}}
\newcommand{\etheo}{\end{theorem}}
\newcommand{\eprop}{\end{proposition}}
\newcommand{\erema}{\end{remark}}
\newcommand{\ecoro}{\end{corollary}}
\newcommand{\elemm}{\end{lemma}}
\newcommand{\eexam}{\end{example}}

\newcommand{\gs}{\bar g}

\title[Standard splittings for conformally stationary spacetimes]{ Existence of standard splittings for conformally stationary spacetimes}

\author[M. A. Javaloyes]{Miguel Angel Javaloyes}
\author[M. S\'anchez]{Miguel S\'anchez}
\email{ma.javaloyes@gmail.com}
\email{sanchezm@ugr.es}
\address{Departamento de Geometr\'{\i}a y Topolog\'{\i}a.
 Facultad de Ciencias, Universidad de Granada.
 Campus Fuentenueva s/n, 18071 Granada, Spain}
\thanks{Both authors partially supported by
Regional J. Andaluc\'{\i}a Grant P06-FQM-01951. MAJ  was also
partially supported by Spanish MEC Grant MTM2007-64504 and MS by
MEC-FEDER Grant MTM2007-60731}
\thanks{  PACS  04.20.Cv, 04.20.Gz, 02.40.Ma}
\thanks{ MSC 2000 : 83C20, 53C50, 53C80 \\
\textbf{Key words:}  Stationary spacetime, global standard
splitting, conformal Killing vector field, causal ladder of
causality, temporal and time functions, smoothability folk
problems.}

\begin{document}

\date{}
\maketitle

\begin{abstract}
 Let $(M,g)$ be a
spacetime which admits a complete timelike conformal Killing
vector field $K$. We prove that $(M,g)$ splits globally as a
standard conformastationary spacetime with respect to $K$ if and
only if $(M,g)$ is distinguishing (and, thus causally continuous).
Causal but non-distinguishing spacetimes with complete stationary
vector fields are also exhibited. For the proof, the recently
solved ``folk problems'' on smoothability of time functions
(moreover, the existence of a {\em
temporal} function) are used.
\end{abstract}



\section{Introduction}

Many physically interesting spacetimes are stationary
or, more generally, conformastationary.
Locally, such a spacetime with a timelike conformal-Killing vector
field $K$ can be written as a {\em standard conformastationary
spacetime} with respect to $K$, i.e.,  a product manifold $M=\R
\times S$ ($\R$ real numbers, $S$ any manifold), where the metric can
be written, under natural identifications, as
 \be \label{st}
g_{(t,x)} 
= \Omega(t,x) \left(-\beta(x) dt^{2}  + 2\omega_x dt +
\gs_x\right), \ee
being 
$\Omega$  a positive function on $M$, and $\gs,\ \beta,\ \omega$,
resp., a Riemannian metric, a positive function and a 1-form, all
on $S$; the vector field $K$ is (locally) identified with
$\partial_t$ (see e.g. \cite{Kr}). The case $\Omega\equiv 1$, or
independent of $t$, corresponds to a {\em standard stationary
spacetime} (notice that,  in general, the function $\beta$ can be
absorbed by the conformal factor $\Omega$).
 Then, a natural question is to wonder when a spacetime admitting
a (necessarily complete) conformastationary $K$ can be written
{\em globally} as above.

The possibility of obtaining  a {\em topological} splitting was
proved by Harris under the assumption of chronology (see the next
section for definitions on causality):

\btheo \label{th} \cite{Ha}. If a spacetime $(M,g)$ admits a
complete stationary vector field $K$ and it is {\em chronological}
then it splits topologically and differentiably as a product $M =
\R \times Q$, where $Q$ is the space of integral curves of $K$,
endowed with a natural manifold structure.

Moreover, for any point $p\in M$ there is a neighborhood $U$ such
that  the projection $\pi: \R\times Q \rightarrow Q$ admits a
local spacelike section $U (\subseteq Q) \rightarrow \R\times Q, x
\mapsto (t(x),x)$ and, then, $\pi^{-1}(U)$ is standard stationary.
\etheo (The result can be easily extended to the
conformastationary case, see Remark \ref{r}). Nevertheless, this
topological splitting does not ensure the existence of the full
metric splitting (\ref{st}), except if one assumes that $\R \times
Q$ admits a global spacelike section ($U=Q$). Our purpose is to
give a full solution to the metric problem, by studying carefully
the involved causality. Concretely:

\btheo \label{t}  Let  $(M,g)$ be a spacetime which admits a
complete conformastationary vector field $K$. Then, it admits a
standard splitting (\ref{st}) if and only if $(M,g)$ is
distinguishing. Moreover, in this case, $(M,g)$ is causally
continuous. \etheo

In the next section we will see how the causality conditions are
natural and optimal for the problem, providing also some examples.
In the last section,  Theorem \ref{t} and related results are
proved. We emphasize that the recent solution (see \cite{BeSa2, BeSa})
on the so--called {\em folk problems on smoothability} is needed
for the proof. Theorem \ref{th} is sketched in the Appendix for completeness.

\section{Causality conditions and Killing fields}

Our conventions and approach will be standard, as in the classical
books \cite{BEE, HE, O, Pe, SW}. Nevertheless, some folk problems
on smoothability (initially suggested  in \cite[p. 1155]{SW-bull})
will be relevant here --at least those concerning time functions.
So, we also recommend the expanded discussion \cite{Sa-bras}
(specially Section 4.6) or the review \cite{MS} (Section 3.8.3 and
Remark 3.77).

$(M,g)$ will denote an $m$-dimensional spacetime $(-,+,\dots,+)$, that is assumed connected
and, when a timelike vector field $K$ is given, $(M,g)$ will be
assumed future time-oriented by $K$. A tangent vector $v\in TM$
will be {\em causal} if it is either timelike ($g(v,v)<0$) or
lightlike ($g(v,v)=0$ and $v\neq 0$). A timelike vector field $K$
is called {\em conformastationary} if it is  conformal-Killing
(any local flow $\Phi_t$ of $K$ is a conformal transformation) 
and {\em stationary} if $K$  is Killing ($\Phi_t$ isometry);
accordingly, $(M,g)$ will be also called conformastationary or
stationary. We begin by showing that causality of
conformastationary spacetimes can be reduced to the  stationary
case.

\blemm Let $K$ be a timelike vector field for $(M,g)$. Then, $K$
is conformal-Killing for $g$ iff $K$ is Killing for the conformal
metric $g^*=-g/g(K,K)$. \elemm

\begin{proof} Obviously, $K$ is conformal-Killing for any metric
conformal to $g$. As $g^*(K,K)$ is constant, the conformal factors
for the local flows of $K$ (which are trivially computable from
the value of $g^*(K,K)$ on the integral curves of $K$) must be
equal to 1, i.e., $K$ is Killing (see \cite[Lemma 2.1]{Sa-trans}
for an alternative reasoning).
\end{proof}

\brema \label{r} {\em As the metric conditions in Theorem \ref{t}
(as well as in Theorem \ref{th}) involve only causality, which is
a conformal invariant, we can replace $g$ by $g^*$, so that we
will assume without loss of generality that $K$ is Killing in what
follows (the reader can also assume $g(K,K)\equiv -1$). }\erema

Let us recall the basic facts about causality  involved in our
main result. $(M,g)$ is {\em chronological} (resp. {\em causal})
if it does not contain any closed timelike (resp. causal) curve.
It is easy to construct a chronological but non-causal stationary
spacetime. Consider in $\R^2$ the vector fields
$X=\partial_t-\partial_x$, $Y=\partial_x$ and let $g_1$ be the
Lorentzian metric such that $X,Y$ are lightlike and $g_1(X,Y)=-1$.
The non-causal cylinder $C=\R\times S^1$ obtained by identifying
$(t,x)\sim (t,x+1)$ admits the projection $K$ of $\partial_t$ as a
stationary vector field.

$(M,g)$ is {\em distinguishing} if $p\ne q$ implies both,
$I^+(p)\neq I^+(q)$ and $I^-(p)\neq I^-(q)$ for any $p,q\in M$. A
causal non-distinguishing  spacetime with a complete stationary
$K$ can be obtained from the previous example as follows. Consider
the product $\R^2\times \R$, $g_2=g_1+dy^2$, choose any irrational
number $a$ and identify $(t,x,y) \sim (t,x,y+1)$, $(t,x,y)\sim
(t,x+1,y+a)$. The quotient spacetime $(\tilde M, \tilde g_2)$ (a
variant of Carter's classical example \cite[Fig. 39, p. 195]{HE})
satisfies the required properties. Notice that Theorem \ref{th} is
applicable to $(\tilde M, \tilde g_2)$ and, in fact, $Q$ is
topologically a torus. Nevertheless, $(\tilde M, \tilde g_2)$ does
not split as standard stationary (apply Corollary \ref{c} below).

The next three steps in the causal ladder are {\em strongly
causal, stably causal} and {\em causally continuous}. Remarkably,
if a complete stationary vector field exists then
``distinguishing'' implies ``causally continuous'' (Proposition
\ref{p2}). This last condition means intuitively that the sets
$I^\pm(p)$ not only characterize $p$ (as in the
distinguishing case) but also vary continuously with $p$. There
are also several characterizations (see e.g. \cite[Sect.
3.9]{MS}), and we will use the following: $(M,g)$ is causally
continuous iff it is distinguishing and {\em reflecting}
i.e., 
$I^+(p) \supseteq I^+(q) \Leftrightarrow I^-(p) \subseteq I^-(q)$ (see \cite[Theorem 3.25, Proposition 3.21]{BEE}).

Moreover, in the proofs, it will be used that causal continuity
implies stable causality. This condition means intuitively that
$(M,g)$  not only  is causal but also remains causal by opening
slightly the timecones. A classical consequence of this definition
(in fact, a {\em folk} characterization,  \cite[Sect.
4.6]{Sa-bras}) is the existence of a {\em time function} $t$,
i.e., a continuous function which is strictly increasing on any
future-directed causal curve. The recent full solution of the folk
problems of smoothability (see \cite{BeSa1,BeSa2, BeSa3}) allows
to characterize stably causal spacetimes as those admitting a {\em
temporal} function $t$, i.e. {\em $t$ is smooth with timelike
past-directed gradient} (in particular, a time function). The
existence of such a function will be essential for our proof.
Notice also that it implies strong causality, i.e., the absence of
``almost closed'' causal curves.

Finally, let us point out  that the two remaining steps in the
ladder of Causality (causal simplicity and global hyperbolicity)
can be characterized in a standard stationary spacetime very
accurately, in terms of Fermat metrics (see \cite{JS}). It is also worth
pointing out that globally hyperbolic spacetimes can be defined as
the {\em causal} ones such that the diamonds $J(p,q)=J^+(p)\cap
J^-(q)$ are compact for all $p,q$ (see \cite{BeSa}). In the previous
example $(\tilde M,\tilde g_2)$ (stationary causal
non-distinguishing), the closures $\overline{J(p,q)}$ are compact,
but some $J(p,q)$ are not. For strongly causal spacetimes, the
compactness of $\overline{J(p,q)}$ suffices to ensure global
hyperbolicity (see \cite[Lemma 4.29]{BEE}, \cite{closure}); in
particular, any standard stationary  spacetime with compact $S$ is
trivially globally hyperbolic, as $J(p,q)$ lies in the compact
region $t^{-1}([t(p), t(q)]$).

\section{Proof of the results} 

 \bprop \label{p2} Let $(M,g)$ be a spacetime with a stationary  $K$.
 If $K$ is complete then $(M,g)$ is reflecting. Thus, if,
additionally, $(M,g)$ is distinguishing, then it is causally
continuous. \eprop
\begin{proof} It is enough to show past reflectivity
$I^+(p) \supseteq I^+(q) \Rightarrow I^-(p) \subseteq I^-(q)$ (the
converse is analogous), and we will  adapt the particular proof in
\cite[Theorem 3.1]{Sa-nonlin}. Take any $p\neq q$ in $M$ and let
$\Phi_t: M\rightarrow M$ the flow of  $K$ at the stage $t\in \R$.
Assuming the first inclusion, it is enough to prove $p_{-\epsilon}
:= \Phi_{-\epsilon}(p) \in I^-(q)$, for all $\epsilon >0$  (notice
that the relation $\ll$ is open and, then, any $p'\ll p$ will lie
also in $I^-(p_{-\epsilon})$ for small $\epsilon$). As
$q_\epsilon:= \Phi_\epsilon(q)  \in I^+(p)$, there exists a
future-directed timelike curve $\gamma$ joining $p$ and
$q_\epsilon$. Then, the future-directed timelike curve
$\gamma_{-\epsilon}:= \Phi_{-\epsilon}\circ \gamma$ connects
$p_{-\epsilon}$ and $q$, as required.
\end{proof}

\noi  Notice that the completeness of $K$ is essential. In fact,
the open subset $M$ of $\LL^2$ obtained by removing a spacelike
semi-axis, that is,  $M=\LL^2\backslash \{(0,x)\in \LL^2: x\leq 0\}$, is not
causally continuous. This example also shows the importance of
completeness in Harris' Theorem \ref{th}, as the space of integral
curves $Q$ is not Hausdorff.

Proposition \ref{p2} will be an essential ingredient of the proof
of Theorem \ref{t}, as it will ensure the existence of a temporal
function. The following consequence shows the consistency of the
results.

\bcoro \label{c}
 Any standard stationary spacetime $(M=\R\times S,g)$ as in
(\ref{st}) is causally continuous, and the projection $t:
M\rightarrow \R$ is a temporal function. \ecoro

\begin{proof} It is enough to prove that $t$ is a temporal function
because in this case the spacetime is distinguishing, and
Proposition \ref{p2} applies. This is a well-known fact, but we
sketch the proof for the sake of completeness. Let
$\gamma(s)=(t(s),x(s))$ be a future-directed differentiable causal
curve. Then, it satisfies the second order
inequality in $\dot t$: $$
g_0(\dot x,\dot
x)+2\omega(\dot x)\dot t-\beta \dot t^2\leq 0.
$$
Moreover, as the Killing field $\partial_t\equiv (1,0)$ is
future-directed, we have that
$$g((1,0),(\dot t,\dot x))=\omega(\dot
x)-\beta \dot t<0.$$ Thus, $\dot t$ must satisfy
$$\beta \dot t\geq \omega(\dot x) +\sqrt{g_0(\dot x,\dot
x)+\omega(\dot x)^2} \geq 0 $$ 
and, in fact,
  $\dot t>0$ (if the equality held,
$\dot x=0$ and $\gamma$ would not be causal). This also implies
that $g(\nabla t, v) > 0$ for all future-directed causal $v$ and,
thus, $\nabla t$ is past-directed timelike, as required.
\end{proof}

The main part of the proof of Theorem \ref{t} is to show that the
following lemma can be applied to any level $t\equiv $ constant of
any temporal function $t$.
 \blemm \label{l0} Let $(M,g)$ be a spacetime with a
complete stationary vector field $K$. If there exists a spacelike
hypersurface $S$ which is crossed exactly once by any integral
curve of $K$ then $(M,g)$ is standard stationary. \elemm

\begin{proof} If $\Phi: \R \times M \rightarrow M$ is the globally
defined flow of $K$, its restriction $\Phi_S: \R\times
S\rightarrow M$ is a diffeomorphism and $(\R\times S, \Phi_S^*g)$
is standard stationary.
\end{proof}

\smallskip

\noindent Lemma \ref{l0} also shows that, as a difference with the
static case (see \cite{SaSe}), the standard stationary splitting cannot be expected to be
unique in any case. Now, we can prove our main result.

\smallskip

\noindent {\em Proof of Theorem \ref{t}}. ($\Rightarrow$). Trivial
from Corollary \ref{c}.

($\Leftarrow$). From Proposition \ref{p2} the spacetime is (in
particular) stably causal  and, thus, it admits a temporal
function $t$. If, say, $0\in $Im$t$, let us see that $S=t^{-1}(0)$
satisfies the hypotheses in Lemma \ref{l0}. Notice that $S$ is
spacelike due to the temporality of $t$. Now, consider the line
bundle $\pi: \R \times Q\rightarrow Q$ onto the manifold of integral
curves (Theorem \ref{th}). The restriction of $\pi$ to $S$
is clearly injective (as $S$ is achronal) and open  (say, as a
consequence of the Theorem of the Invariance of the Domain). So,
it is enough to prove that $\pi(S)$ is closed.

Let $\{z_n= (\hat t_n, \hat x_n)\}_n$ be a sequence in $S \subset
\R \times Q$ with $\{\hat x_n\}_n$ converging to some $\hat
x_\infty \in Q$. As $S$ is closed, if we assume by contradiction
that $\hat x_\infty \not\in \pi(S)$, then necessarily $\{\hat
t_n\}_n$ diverges up to a subsequence. We will assume $\{\hat
t_n\}_n \rightarrow +\infty$, as the case $\{\hat t_n\}_n
\rightarrow -\infty$ is analogous.

Let $\hat V$ be some neighborhood of $\hat x_\infty$ such that
$\pi^{-1}(\hat V)= \R\times \hat V$ is standard stationary.
Choosing such a standard splitting, we can write $\pi^{-1}(\hat
V)= \R\times V$ for some spacelike hypersurface $V\subset
\pi^{-1}(\hat V)$ and, consistently:  $\{z_n= (t_n,  x_n)\}_n$
with $\pi(z_n)=\hat x_n$, and (as $V$ is obtained as a section on $\hat V$, according to Th. 1.1) $\{x_n\}_n \rightarrow x_\infty \in V$
with $\pi((0,x_\infty)) = \hat x_\infty$.

Let $W\subset V$ be a compact neighborhood of $x_\infty$. Notice
that there is a constant $T_W>0$ such that $ [t_0 + T_W, +\infty)
\times W \subset I^+(t_0,x_0)$ for any $(t_0,x_0)\in \pi^{-1}(W)$.
In fact, consider the future Fermat arrival function $T:V\times V
\rightarrow [0,+\infty)$, i.e. $T(x_1,x_2)$ is the infimum of the
$t\geq 0$ such that $(x_1,0) \ll (x_2,t)$. As $T$ is continuous
(see \cite[Prop. 2.2]{Sa-pams}) one can take any $T_W\geq T(W,W)$.

Thus, fix some $n_0\in \N$ so that $x_{n_0} \in W$. For large $n$,
$(t_n,x_n) \in  [t_{n_0} + T_W, +\infty) \times W \subset
I^+(t_{n_0}, x_{n_0})$, in contradiction with the achronality of
$S$.\qed
\section{Appendix}


As commented above, Harris' Theorem \ref{th} (which is stated in a
somewhat more general form in \cite{Ha}) can be extended to the
conformal case by using Remark \ref{r}. For the sake of
completeness, we sketch the proof, readapting Harris' arguments.
First, the following general result  is needed \cite[Theorem
2]{Ha}:

\btheo \label{th2} Let $(M,g)$ be a chronological spacetime with a
complete timelike vector field $K$. Then $M$ is naturally a
principal line bundle over the space $Q$ of integral curves of
$K$, and $Q$ is a near-manifold. \etheo \noindent Here, a {\em
near-manifold} is a topological space which satisfies all the
axioms of a smooth manifold (including paracompactness)  except at
most to be Hausdorff. The  proof is subtle, and uses  general
properties of actions of  groups by Palais (see \cite{Pa}).
Now, Theorem
\ref{th} follows directly from:

\blemm In the hypotheses of Theorem \ref{th2}, if $K$ is Killing
then $Q$ is Hausdorff. \elemm

\noindent {\em Proof.} Notice first that $K$ is also Killing for
the Riemannian metric $g_R$ obtained by reversing the sign of $K$,
i.e.:
$$
g_R(v,v)=g(v,v)-\frac{2}{g(K,K)}g(K,v)^2, \quad \forall v\in TM,
$$
and the problem becomes purely Riemannian. Assume that the
projections of $x,y\in M$ on $Q$ are different ($\pi(x)\neq
\pi(y)$) and cannot be separated by disjoint neighborhoods. Then,
for any $\epsilon>0$ the $g_R$-balls $B(x,\epsilon/2)$,
$B(y,\epsilon/2)$ have non-disjoint projections, and there exists
some $t_\epsilon\in \R$ such that
$\Phi_{t_\epsilon}(B(x,\epsilon/2)) \cap B(y,\epsilon/2)\neq
\emptyset$. That is, there exists some integral curve $\gamma$ of
$K$ such that $\gamma(0)\in B(x,\epsilon/2)$ and
$\gamma(t_\epsilon)\in B(y,\epsilon/2)$. Therefore, as
$\Phi_{t_\epsilon}(B(x,\epsilon/2))=B(\phi_{t_\epsilon}(x),\epsilon/2)$,
we have that $y\in B(\phi_{t_\epsilon}(x),\epsilon)$, and this
implies  that $\pi(y)\in \pi (B(\Phi_{t_\epsilon}(x),\epsilon))=
\pi(B(x,\epsilon))$ for all $\epsilon>0$. As a consequence, $Q$ is
not a $T_1$ topological space, in contradiction with Theorem
\ref{th2}. \qed

\end{document}